\documentclass[onecolumn]{article}

 \evensidemargin \oddsidemargin

\usepackage{palatino,epsfig,amssymb,color}
\usepackage[fleqn]{amsmath}
\usepackage{natbib}

\newtheorem{lemma}{Lemma}
\newtheorem{theorem}{Theorem}

\def\QED{~\rule[-1pt]{5pt}{5pt}\par}
\newenvironment{proof}{{\em Proof.}}{{ \hfill \QED}\medskip}

\DeclareMathAlphabet{\matheur}{U}{eur}{m}{n}
\DeclareMathAlphabet{\matheurb}{U}{eur}{b}{n}
\DeclareMathAlphabet{\matheus}{U}{eus}{m}{n}
\DeclareMathAlphabet{\matheuf}{U}{euf}{m}{n}

\newcommand{\hs}{\hspace{3mm}}

\newcommand{\IR}{\mathbb{R}}
\newcommand{\IL}{\mathbb{L}}
\newcommand{\IH}{\mathbb{H}}
\newcommand{\IE}{\mathbb{E}}
\newcommand{\IS}{\mathbb{S}}

\newcommand{\eps}{\varepsilon}

\newenvironment{mat}{\left[\begin{array}}{\end{array}\right]}

\DeclareMathAlphabet{\matheur}{U}{eur}{m}{n}
\DeclareMathAlphabet{\matheurb}{U}{eur}{b}{n}
\DeclareMathAlphabet{\matheus}{U}{eus}{m}{n}
\DeclareMathAlphabet{\matheuf}{U}{euf}{m}{n}

\renewcommand{\varrho}{\rho}
\newcommand{\rhol}{{\varrho}}

\newcommand{\RHinf}{\IR \IH_\infty}
\newcommand{\RLinf}{\IR \IL_\infty}

\newcommand{\Rset}{\mathbb{R}}
\newtheorem{prop}{Proposition}

\usepackage[skip=3mm]{parskip}

\usepackage[normalem]{ulem}

\title{ %
Instability Margin Analysis for Parametrized LTI Systems 
with Application to Repressilator 
}

\author{Shinji Hara, Tetsuya Iwasaki, Yutaka Hori%
\thanks{S. Hara is with Systems and Control Engineering, Tokyo Institute of Technology, 2-12-1 Ohokayama, Meguro-ku, Tokyo, Japan. T. Iwasaki is with Mechanical and Aerospace Engineering, University of California Los Angels, 420 Westwood Plaza, Los Angeles, CA 90095. 
Y. Hori is with Applied Physics and Physico-Informatics, Keio University, 3-14-1 Hiyoshi, Kohoku-ku, Yokohama, Kanagawa 223-8522, Japan. \newline
Corresponding author T. Iwasaki. {Tel. +1-310-206-2533. Fax. +1-310-206-2302.}}%
}

\date{}

\makeatletter
\def\thanks#1{\protected@xdef\@thanks{\@thanks
\protect\footnotetext{#1}}}
\makeatother

\begin{document}
\maketitle

\begin{abstract} 
This paper is concerned with a robust instability analysis for the single-input-single-output unstable 
linear time-invariant (LTI) system under dynamic perturbations. The nominal system itself is possibly 
perturbed by the static gain of the uncertainty, which would be the case when a nonlinear uncertain 
system is linearized around an equilibrium point. 
We define the robust instability radius as the smallest $H_\infty$ norm of the stable linear perturbation that 
stabilizes the nominal system. There are two main theoretical results: one is on a partial characterization 
of unperturbed nominal systems for which the robust instability radius can be calculated exactly, 
and the other is a numerically tractable procedure for calculating the exact robust instability radius
for nominal systems parametrized by a perturbation parameter. The results are applied to the repressilator 
in synthetic biology, where hyperbolic instability of a unique equilibrium guarantees the persistence 
of oscillation phenomena in the global sense, and the effectiveness of our linear robust instability analysis 
is confirmed by numerical simulations. 
\vspace*{-3mm}
\end{abstract}
\section{Introduction}
\label{sec:Intro}

\vspace*{-2mm}

Feedback control to
maintain non-equilibrium state such as oscillation has been recognized as an important
design problem for engineering applications including robotic locomotion 
\citep{grizzle:01,wu:21}. Such non-equilibrium state may be robustly 
maintained if every equilibrium point is hyperbolically unstable
\citep{pogromsky:99}. This fact motivates robust instability analysis of an 
equilibrium point subject to perturbations. 
For linear systems, analysis of robust instability is equivalent to 
finding the minimum norm stable controller that stabilizes a given unstable plant, 
which is known to be extremely difficult due to the requirements of the strong 
stabilization \citep{youla:74} and the norm constraint on the controller. 
The analysis is further complicated by the fact
that equilibrium points may change due to perturbations in nonlinear dynamical systems.
When the equilibrium is perturbed, the linearized dynamics would be 
altered, and hence an analysis of a fixed linearized system no longer 
characterizes the robustness property of the equilibrium point.  
This issue has been pointed out in the context of a robust stability 
analysis for Lur'e type nonlinear systems \citep{wada:98,wada:00}, 
as well as in a robust bifurcation analysis \citep{inoue:15}. 
Thus, we need to develop a theory to address this issue properly with a
general framework to lay a foundation for the linear robust instability theory. 

\vspace*{-2mm}

In this paper, we formally define the robust instability radius (RIR) for
single-input-single-output (SISO) unstable linear time invariant 
(LTI) systems subject to dynamic perturbations \citep{inoue:13ecc}
in a manner analogous to the classical robust stability radius 
\citep{hinrichsen:86a}.  A key technical result shows that the RIR analysis 
reduces to a marginal stabilization problem, leading to two conditions
under which the exact RIR is given as the inverse of the 
static or peak gain of the system.  
We will then extend our analysis to rigorously take account of the possible
change of the nominal linear dynamics caused by the perturbation.  
The main theoretical result of this part leads to a computationally tractable 
procedure to find the exact RIR for a class of parametrized LTI 
systems. Finally, the effectiveness of the theoretical results 
will be demonstrated through numerical simulations by an application to the 
repressilator \citep{elowitz:00}. 

\vspace*{-2mm}

Our approach builds on the preliminary result \citep{hara:20}, which formalized 
the robust instability analysis problem for a fixed LTI system by 
introducing a notion of the RIR.
The contributions of the present paper beyond \citep{hara:20} include
theoretical justification of the marginal stabilization approach, a characterization
of third order systems for which the RIR can be found exactly, and an extension
to the parametrized LTI systems to account for the change of the nominal dynamics
due to the perturbation. 

\vspace*{-2mm}

The remainder of this paper is organized as follows. 
Sections~\ref{sec:RIR} and~\ref{sec:InstabilityMargin} are  devoted to the 
analyses of the RIR for fixed and parametrized LTI systems, respectively. 
The effectiveness of the theoretical results is confirmed by an application 
to the repressilator model in Section~\ref{sec:Repressilator}. 
Section~\ref{sec:Concl} summarizes the contributions of this paper 
and addresses some future research directions.

\vspace*{-2mm}

We use the following notation. 
The set of real numbers is denoted by $\Rset$. 
$\Re(s)$ denotes the real part of a complex number $s$. 
The set of real rational functions bounded on $j\Rset$ is 
denoted by $\RLinf$, and its stable subset by $\RHinf$.
The norms in these linear spaces are denoted by
$\| \cdot \|_{L_\infty}$ and $\| \cdot \|_{H_\infty}$, respectively.
The open left and right half complex planes are abbreviated as OLHP and ORHP, respectively. 

\vspace*{-2mm}

\section{Robust Instability Radius for LTI Systems}
\label{sec:RIR}

\vspace*{-2mm}

This section is devoted to the analysis of the robust instability radius
(RIR) for a given unstable transfer function $g(s) \in \RLinf$. 
We will provide two classes of $g(s)$ for which the RIR can be characterized exactly. 

\subsection{Definition and Preliminary Results on RIR}
\label{subsec:DefRIR}

\vspace*{-2mm}

Our target system is an unstable system represented by the transfer function $g(s)$ 
which has no poles on the imaginary axis, i.e., $g(s) \in \RLinf$.   
Given such $g(s)$, we introduce a set denoted by $\IS(g)$ as follows: 
It is the set of $\RHinf$ functions $\delta(s)$ that internally stabilizes $g(s)$ with positive feedback, 
that is, $\IS(g)$ is defined as
\vspace*{-3mm} 
\begin{eqnarray}
 \IS(g)  &:=& \Big\{ ~ \delta(s) \in \RHinf:~   \nonumber \\ 
\vspace*{-1mm}
&& 
\begin{array}{l}
\delta(s)g(s)=1 ~ \Rightarrow ~ \Re(s)<0 \vspace*{-1mm} \\ 
\delta(s)=0, \Re(s)>0 ~ \Rightarrow ~ |g(s)| < \infty 
\end{array}
\Big\} . 
\vspace*{-2mm}
\end{eqnarray}
The first condition of $\IS(g)$ means that the characteristic roots of the
positive feedback connection of $\delta(s)$ and $g(s)$ are in the OLHP, 
and the second one implies that $\delta(s)$ and $g(s)$ have no unstable pole/zero 
cancellation for the internal stability. 
In this sense, "{\em stabilization}" in this paper means "{\em internal stabilization}." 

\vspace*{-2mm}

Let us first define the robust instability radius (RIR) for 
a given $g(s)\in\RLinf$, which will be useful for later developments. 
The RIR for $g(s)$, denoted 
by $\rho_*$, is defined to be the magnitude of the smallest perturbation
that stabilizes $g(s)$, {\it i.e.,} 
\begin{equation} \label{eq:RIR0}
\vspace*{-2mm}
\rho_*:=\inf_{\delta\in\IS(g)} ~ \|\delta\|_{H_\infty}.
\vspace*{-2mm}
\end{equation} 
It is clear from the condition for the strong stabilizability in \citep{youla:74} 
that $\IS(g)$ is nonempty and hence $\rho_*$ for 
$g(s)$ is finite if and only if the Parity Interlacing Property (PIP) 
is satisfied, {\it i.e.,} the number of unstable real poles of $g(s)$ between any pair of real zeros 
in the closed right half complex plane (including zero at $\infty$) is even.

\vspace*{-2mm}

Some lower bounds of $\rho_*$ are known from the literature as follows.

\vspace*{-2mm}

\begin{lemma} \rm \citep{inoue:13ecc,inoue:13cdc,hara:20} \;  
\label{prop:lbub}
Let $g(s)\in\RLinf$ be given. Suppose $g(s)$ is strictly proper and unstable. 
Then 
\begin{equation} \label{rhoP}
\vspace*{-3mm}
 \rho_*\geq ~
 \rhol_p:=1/\|g\|_{L_\infty} , \hs \|g\|_{L_\infty} := \sup_{\omega \in \IR} |g(j\omega)| . 
 \vspace*{-3mm}
\end{equation} 
Moreover, if $g(s)$ has an odd number of unstable poles (including multiplicities) then we have 
\begin{equation} \label{lbo}
\vspace*{-3mm}
 \rho_* \geq \rhol_o := 1/|g(0)|.
\vspace*{-3mm}
\end{equation}
\end{lemma}

\vspace*{-2mm}

The lower bounds given above can readily be calculated,
immediately giving an estimate for the RIR.  It will turn out
later that each of these bounds is tight for a certain case, providing the 
exact value of the RIR $\rho_*$. 
To that end, we will develop an approach for characterizing an upper bound in a tractable manner,
and provide conditions under which the upper bound coincides with one of the lower bounds
$\rhol_o$ and $\rhol_p$. 
These are addressed in the following subsections. 

\vspace*{-2mm}

\subsection{Upper bound via marginal stabilization}

\vspace*{-2mm}

An upper bound is obtained as $\|\delta\|_{H_\infty}$ if a stable 
stabilizing perturbation $\delta(s)$ is found.  
Since the closed-loop poles for such perturbation are in the {\em open} left half plane, a scaled
perturbation $(1-\eps)\delta(s)$ is also stabilizing for sufficiently small
$\eps>0$, and has smaller norm $(1-\eps)\|\delta\|_{H_\infty}$, yielding a
better (smaller) upper bound.  
This observation leads to the fact that the best (least norm) upper bound is necessarily obtained 
from a perturbation that marginally stabilizes the closed-loop system.  
Hence, we may focus on the search for a marginally stabilizing, stable perturbation $\delta_o(s)$. 
However, such $\delta_o(s)$ does not necessarily give an upper bound 
since a slight perturbation of $\delta_o(s)$ may not be able to (strictly) 
stabilize the closed-loop system in general.  
The following result shows that an upper bound can always be obtained if marginal stability is
achieved with a single mode on the imaginary axis.

\vspace*{-2mm}

\begin{prop} \label{PropS2}
Consider real-rational transfer functions $g(s)$ and $\delta_o(s)$ 
having no unstable pole/zero cancellation between them, 
where the former is strictly proper and the latter is proper and 
stable (possibly a real constant). 
Suppose $\delta_o(s)$ marginally stabilizes $g(s)$ 
with the closed-loop characteristic roots of $\delta_o(s)g(s)=1$ all in the OLHP
except for either a pole at the origin or a pair of complex conjugate poles on the imaginary axis. 
Then, for almost\footnote{
This means that an arbitrarily chosen $\delta_1(s)$ may or may not work to stabilize, 
but when it does not work, a slight modification of it can always make it work.}
any proper stable transfer function $\delta_1(s)$, 
there exists $\eps\in\IR$ of arbitrarily 
small magnitude $|\eps|$ such that the positive feedback with 
$\delta_\eps(s):=\delta_o(s)+\eps\delta_1(s)$ internally stabilizes $g(s)$. 
\end{prop}

\vspace*{-2mm}
See Appendix~\ref{subsec:ProofProp3} for a proof of the proposition. 

\vspace*{-2mm}

When a stable perturbation $\delta_o(s)$ achieves marginal stability
with multiple modes on the imaginary axis (which is a rather rare occasion), 
every slight modification of $\delta_o(s)$ may move at least one purely 
imaginary pole into the right half plane. In this case, 
$\|\delta_o\|_{H_\infty}$ is not an upper bound on the RIR. However, 
Proposition~\ref{PropS2} shows that, when there is a single mode on the 
imaginary axis (which is generically expected), almost every perturbation 
of $\delta_o(s)$ moves the imaginary pole(s) in a direction transverse to the 
imaginary axis, and hence it is always possible to strictly stabilize the 
closed-loop system. Therefore, an upper bound can be obtained by searching 
for such $\delta_o(s)$. 

\vspace*{-2mm}

Given the extreme difficulty of strong stabilization with the minimum
norm controller, Proposition~\ref{PropS2} is significant because (a) the 
search for such marginally stabilizing $\delta_o(s)$ can be performed 
systematically by restricting our attention to some specific class of 
transfer functions, and (b) this approach is suitable for the search
for the minimum norm, marginally stabilizing perturbation $\delta_o(s)$. 
These claims are explained in the next section.

\vspace*{-2mm}

\subsection{Search for marginally stabilizing perturbation}
\label{subsec:ExactAnalysis}

\vspace*{-2mm}

First note that marginal stability requires that $\delta_o(s)$ be chosen to satisfy
\begin{equation} \label{dp}
\vspace*{-3mm}
\delta_o(j\omega_c)=\delta_c := 1/g(j\omega_c), 
\vspace*{-3mm}
\end{equation}
at a critical frequency $\omega_c\geq0$, so that $s=j\omega_c$ is a closed-loop pole. 
If we parametrize a class of perturbations, then $\delta_o(s)$ satisfying
(\ref{dp}) may be determined for each $\omega_c\in\IR$, 
and an upper bound $\|\delta_o\|_{H_\infty}$ on the RIR is obtained when 
the resulting closed-loop poles (i.e. roots of $\delta_o(s)g(s)=1$) are all 
in the OLHP except for $s=\pm j\omega_c$ 
(let us call this property $\omega_c$-stability).

A reasonable candidate for the class of $\delta_o(s)$ is the
set of all-pass transfer functions \citep{hara:20}. 
An advantage of using all-pass functions is that the least upper bound
on the RIR is obtained for a given $\omega_c$ since the gap in
$\|\delta_o\|_{H_\infty}\geq|\delta_o(j\omega_c)|$ is eliminated regardless
of the value of $\omega_c$. 
The simplest choice is the first (or zeroth) order all-pass function expressed as  
\begin{equation} \label{do}
\vspace*{-3mm}
 \delta_o(s)=b\cdot\frac{s-a}{s+a},  \hs  (a  \geq 0) .  
\vspace*{-3mm}
\end{equation} 
There are two requirements for choosing the parameters $(a,b)$ in $\delta_o(s)$. 
One is (\ref{dp}) and the other is $a \geq 0$ to assure stability
\footnote{$a=0$ is allowed since $\delta_o(s)$ becomes constant.} 
of $\delta_o(s)$.
A simple calculation leads to $|b|=|\delta_c|$ and 
$\angle(j\omega_c-a)-\angle(j\omega_c+a)+\angle(b)=\angle(\delta_c)$,
which gives the proper choice of $(a,b)$ as follows:
\begin{equation} \label{eq:ab} 
\vspace*{-2mm}
\begin{array}{lll}
a=\omega_c\tan\varphi, & b=|\delta_c|, & (0\leq\varphi<\pi/2), \\
a=\omega_c\tan(\varphi+\pi/2), & b=-|\delta_c|, & (-\pi/2\leq\varphi<0), 
\end{array} 
\vspace*{-2mm}
\end{equation}
where $\varphi:=\angle\delta_c/2$.  
Thus, for a given $\omega_c\in\IR$, the stable first order all-pass function (\ref{do}) is uniquely determined.  
The gain $\|\delta_o\|_{H_\infty}$ is then an upper bound on the RIR if 
the $\omega_c$-stability is achieved by $\delta_o(s)$.  
Sweeping over $\omega_c\in\IR$, the least upper bound within this framework can be calculated. 

\vspace*{-2mm}

\subsection{Simple classes of $g(s)$ for which the exact RIR can be
analytically characterized}
\label{subsec:ExactClass}

\vspace*{-2mm}
In view of Lemma~\ref{prop:lbub}, important cases occur 
when the $\omega_c$-stability is achieved at $\omega_c=0$ or $\omega_p$, 
where $\omega_p$ is the peak frequency at which 
$|g(j\omega_p)|=\|g\|_{L_\infty}$ holds.  
In these cases, the exact values of the RIR is given by $\rho_*=1/|g(j\omega_c)|$ 
because the upper and lower bounds coincide. 
In particular, we have the following:
\vspace*{-2mm}
\begin{itemize}
\item
$\rho_*=\varrho_o$ if $g(s)$ satisfies {\em Condition 1}:  $g(s)\in\RLinf$ has 
an odd number of poles (including multiplicities) in the ORHP, and the constant 
perturbation $\delta_o(s)=1/g(j\omega_c)$ achieves 
the $\omega_c$-stability with $\omega_c=0$. 
\item 
$\rho_*=\varrho_p$
if $g(s)$ satisfies {\em Condition 2}:  $g(s)\in\RLinf$ is unstable,  
and $\delta_o(s)$ in (\ref{do}) with (\ref{eq:ab}) achieves 
$\omega_c$-stability at $\omega_c=\omega_p$.
\end{itemize}

While it is easy to construct $\delta_o(s)$ numerically for a given $g(s)$ and 
check $\omega_c$-stability of the closed-loop system, it remains open to 
fully characterize the class of $g(s)$ which can be $\omega_c$-stabilized by 
a first-order all-pass function.  
Here we present a subclass of third order systems for which the above idea works. 
This class covers the simplest model of 
the repressilator in synthetic biology as seen in Section~\ref{sec:Repressilator}. 

\vspace*{-2mm}

\begin{prop} \label{coro:3rd}
Consider the third order transfer function represented by 
\begin{equation} \label{g3rd}
g(s) = \frac{\zeta s - k}{s^3+ps^2+qs+\ell}, \hs k \neq 0
\end{equation}
(i) $g(s)$ satisfies Condition 1 if and only if
\begin{equation} \label{c22}
p>0, \hs \ell<0, \hs q + \zeta \ell/k > 0,
\end{equation}
which implies $\rho_*=\varrho_o:=1/|g(0)|$. \\
(ii) $g(s)$ satisfies Condition 2 if
\begin{equation} \label{c31} 
p>0, \hs \ell > pq, \hs q^2 < 2p\ell,
\end{equation}
which implies $\rho_*=\varrho_p:=1/\|g\|_{L_\infty}$.
\end{prop}
\vspace*{-2mm}
See Appendix~\ref{subsec:ProofsCorollaries} for a proof of the proposition.
\vspace*{-2mm}

There are three remarks on the class of $g(s)$ of the form (\ref{g3rd}) satisfying (\ref{c31}):
(i) The requirement of $\ell > pq$ is necessary and sufficient for $g(s)$ 
to have two unstable poles, provided $p>0$ and $\ell>0$, which are  
guaranteed by the first and third inequalities in (\ref{c31}). 
(ii) The requirement of $q^2 < 2p\ell$ is a sufficient condition 
for the infinity norm of $g(s)$ to be attained at a non-zero frequency. 
(iii)  A class of third order systems represented by 
$g(s)=k/((s+\alpha)(s^2 -\beta s + \gamma^2))$ with $0<\beta<\gamma<\alpha$ 
satisfies (\ref{c31}) and hence Condition 2, which was numerically 
verified earlier in \citep{hara:20}.

\vspace*{-2mm}

\section{RIR for Parametrized LTI Systems} 
\label{sec:InstabilityMargin}

\vspace*{-2mm}

Here we derive a general theoretical result to provide a computationally tractable 
method for characterizing the exact robust instability radius $\mu_*$ for 
a class of parameterized linear systems based on the instability radius analysis 
in the previous section. 

\subsection{Definition of RIR $\mu_*$}
\label{subsec:ProblemNonlinear}

\vspace*{-2mm}

We consider a family of SISO transfer functions $g_e(s)$ parameterized by 
$e\in\Rset$, where we assume that $g_e(s)$ has at least one pole in the 
ORHP and no poles on the imaginary axis for all $e\in\, \IE := (e_-, e_+)$ 
which includes the origin, i.e., $e_- < 0 < e_+$.  
For each $e\in\IE$, let $\Delta_e$ be the set of all perturbations $\delta(s)\in\RHinf$ 
that stabilizes $g_e(s)$ and satisfies $\delta(0)=e$, 
{\it i.e.,}
\begin{equation} \label{Deltae} 
\vspace*{-2mm}
\Delta_e:=\{\delta(s)\in\IS(g_e):~ \delta(0)=e ~ \} . 
\vspace*{-2mm}
\end{equation}
The objective is to calculate {\em the robust instability radius} $\mu_*$ for $g_e(s)$
defined by 
\begin{equation} \label{re}
\vspace*{-2mm}
\mu_* := \inf_{e\in\IE}~\mu(e), \hs
\mu(e) := \inf_{\delta(s)\in\Delta_e}\|\delta\|_{H_\infty},
\vspace*{-2mm}
\end{equation}
where we define $\mu(e) := \infty$ if $\Delta_e$ is empty and 
$\mu_* := \infty$ if $\Delta_e$ is empty for all $e \in \IE$.
Note that $\rho_*$ is identical to $\mu(e)$ for $g_e(s)=g(s)$ except for the 
absence of the constraint on the static gain $\delta(0)=e$. 
The idea is that $g_e(s)$ is the system obtained by linearization of
a nonlinear system around an equilibrium point, perturbed by the uncertainty 
$\delta(s)$ with static gain $e$.   

As an example, let us consider a dynamically perturbed 
FitzHugh-Nagumo (FHN) neuron model presented in  
\citep{hara:20}. We will show how to derive the 
corresponding $g_e(s)$ based on the uncertain FHN model %
\[
\begin{array}{l}
c \dot{v} = \psi(v) - (1+\delta(s))w, \hs \psi(v):=v-v^3/3, \\
\tau \dot{w} = v+\alpha-\beta w,
\end{array} 
\]
with positive constant parameters $c$, $\tau$, $\alpha$, and $\beta$, 
where the term $\tilde{w} := (1+\delta(s))w$ represents the 
dynamically perturbed $w$ with uncertainty $\delta(s)$. 
Let $(\bar v,\bar w)$ be an equilibrium point, i.e., 
\[
\psi(\bar v)=(1+e)\bar w, \hs \bar v=\beta \bar w-\alpha 
\]
hold, where $e:=\delta(0)$. It can be verified that the equilibrium is 
unique if $1+e>\beta$. Linearizing the system around $(\bar{v},\bar{w})$, 
the characteristic equation is given by $1=\delta(s)g_e(s)$ with 
\[ 
g_e(s) := 1/\Big(c\tau s^2+(\beta c-\tau\gamma)s+1-\beta\gamma\Big), 
\]
where $\gamma:=\psi'(\bar v)=1-\bar{v}^2$. Note that $g_e(s)$ 
depends on $e$ through $\bar v$ since $\gamma$
is a function of $e$. The perturbation $\delta(s)$ stabilizes the
equilibrium when it stabilizes $g_e(s)$ and has the consistent static
gain $\delta(0)=e$.

The term RIR refers to both $\rho_*$ for a fixed LTI system and 
$\mu_*$ for parametrized LTI systems. While the mathematical definitions 
of $\rho_*$ and $\mu_*$ are different, both represent the smallest magnitude 
of perturbations that stabilize the underlying system (or equilibrium). 
In the special case where $g_e(s)$ is independent of $e$ (i.e. the 
equilibrium does not move by perturbation $\delta(s)$) and $\IE=\IR$, 
the RIR $\mu_*$ reduces to $\rho_*$.
The final goal of this paper is provide a computable characterization for the 
robust instability radius $\mu_*$, which is the magnitude of the smallest perturbation 
$\delta(s)$ that stabilizes the equilibrium point of the nonlinear system.

\vspace*{-2mm}

\subsection{Lemmas for Exact Analysis}
\label{subsec:Lemmas}

\vspace*{-2mm}

This section presents three lemmas as preliminaries.  
All the proofs are given in Appendix~\ref{subsec:ProofsLemmas}. 
Let us first examine the relationship between $\mu_*$ and $\rho_*$.  
The minimum value of $\mu(e)$ over $e \in \IE$ is $\mu_*$ as seen in (\ref{re}). 
A simple observation shows that $\mu(e)$ is closely related to the RIR for linear system $g_e(s)$,
denoted by $\rho_*(e)$:
\begin{equation} \label{rhoe}
\rho_*(e) := \inf_{\delta \in \IS(g_e)} \|\delta\|_{H_\infty} .
\end{equation}
In particular, $\rho_*(e)$ is a lower bound of $\mu(e)$.   

\begin{lemma} \label{lem:lbre}
For each $e\in\IE$, we have 
\[
\mu(e) \geq |e|, \hs \mu(e) \geq \rho_*(e).
\]
\end{lemma}

\vspace*{-2mm}

Based on the results for the linear case, we expect that 
\[
\varrho_o(e):=1/|g_e(0)|, \hs
\varrho_p(e):=1/\|g_e\|_{L_\infty},
\]
may play an important role in characterizing $\mu_*$. 
In view of the results in the previous section, $\rho_*(e)$ is exactly characterized by
$\varrho_o(e)$ or $\varrho_p(e)$, provided $g_e(s)$ satisfies Condition 1 or 2, respectively. 
Hence, the key for characterizing $\mu_*$ is to obtain the condition under which
the lower bound $\rho_*(e)$ on $\mu(e)$ is tight, in which case,
$\mu_*$ is given by the infimum of $\rho_*(e)$ over $e\in\IE$.
A technical difficulty is that when a stabilizing perturbation $\delta(s)$ is found for $g_e(s)$, 
it is likely that $\delta(0)=e$ is violated and hence such $\delta(s)$ cannot be used
for the calculation of $\mu(e)$ in (\ref{re}).  
The following result is useful for adjusting the static gain of $\delta(s)$ by a high pass filter 
while preserving the stabilizing property.

\begin{lemma} \label{lem:hpf}
Let $\gamma\in\IR$ and scalar-valued, strictly proper, real-rational transfer 
function $\ell(s)$ be given. Suppose $\ell(s)$ has an even number of poles  
(including multiplicities) in the ORHP and no poles on the imaginary axis, 
all the roots of $1=\ell(s)$ are in the OLHP, and
\begin{equation} \label{sgc}
|\gamma|<1, \hs \|\gamma \ell\|_{L_\infty}<1
\end{equation}
hold. Then, for sufficiently small $\xi>0$, all the roots of
\begin{equation} \label{chefl}
1=f(s)\ell(s), \hs f(s):=\frac{s+\xi\gamma}{s+\xi}
\end{equation}
are in the OLHP.
\end{lemma}

Using Lemma~\ref{lem:hpf} and the ideas from the RIR analysis, 
we can characterize $\mu(e)$ as follows.

\begin{lemma} \label{lem:re} 
Fix $e \in \IR$, let a strictly proper transfer function $g_e(s)$ be given, 
and consider $\Delta_e$ and $\mu(e)$ in (\ref{Deltae}) and (\ref{re}), respectively.  
Suppose $|e| < \varrho_p(e) := 1/\|g_e\|_{L_\infty}$ holds
and $g_e(s)$ satisfies the following conditions: 
\vspace*{-2mm}
\begin{itemize} 
\item[(a)] $g_e(s)$ satisfies Condition 2. 
\item[(b)] $g_e(s)$ has 
a nonzero even number of poles (including multiplicities) in the ORHP. 
\end{itemize}
\vspace*{-2mm}
Then $\mu(e) = \varrho_p(e)$ holds.
\end{lemma}

Condition {\it (a)} in Lemma~\ref{lem:re} guarantees that $\varrho_p(e)$ is the exact 
linear RIR for $g_e(s)$. 
However, the smallest perturbation $\delta(s)$ does not move the equilibrium 
to a point at which $g_e(s)$ is the corresponding linearization unless $\delta(0)=e$. 
Condition {\it (b)} allows for the use of Lemma~\ref{lem:hpf} to adjust the static gain 
of the perturbation so that $\delta(s)$ with a high pass filter 
has the static gain $e$ and thus stabilization of $g_e(s)$ corresponds to 
stabilization of the original equilibrium point of the nonlinear system. 
It can be shown that conditions {\it (a)} and {\it (b)} hold for the class of third order 
systems with (\ref{c31}) in Proposition~\ref{coro:3rd}. 

The static gain adjustment does not work for $g_e(s)$ satisfying Condition 1
since the number of poles in the ORHP is odd, in which case the perturbation with the 
high pass filter destabilizes $g_e(s)$.  
Hence, $\mu(e) = \varrho_o(e)$ does not hold in general under Condition 1. 
The property $\mu(e) \geq \rho_*(e) = \varrho_o(e)$ 
is still useful for obtaining lower and upper bounds on $\mu_*$, but does not seem
to yield an exact characterization of $\mu_*$. 
Therefore, we will focus on the case where $g_e(s)$ satisfies the conditions in Lemma~\ref{lem:re} 
in the next subsection.

\subsection{Exact RIR Analysis}
\label{subsec:ExactIM}

\vspace*{-2mm}

Lemma~\ref{lem:re} characterizes $\mu(e)$ only when $e$ satisfies $|e| < \varrho_p(e)$, 
and does not cover all possible values of $e \in \IE$.
However, it turns out that the minimum of $\mu(e)$ over $e\in\IE$ occurs
within the subset of $\IE$ where $|e| < \varrho_p(e)$ holds, 
and hence we have a computable description of $\mu_*$ as stated in the following theorem.

\vspace*{-2mm}

\begin{theorem} \label{theorem:nonlinear} 
Consider the parametrized LTI system $g_e(s)$ with $e\in\IE\subset\Rset$,
where $0\in\IE$.  
Let $\IE_* \subset \IE$ be the largest interval such that 
$|e| < \varrho_p(e)$ holds for $e\in\IE_*$.  
Then
\begin{equation} \label{eq:R*} 
\mu_* = \inf_{e \in \IE_*} \varrho_p(e)
\end{equation}
holds, provided $g_e(s)$ satisfies conditions {\it (a)} and {\it (b)} in Lemma~\ref{lem:re}
for all $e\in\IE_*$. 
\end{theorem}
\begin{proof}
Let $R_{\rm in}$ be the infimum of $\mu(e)$ over $e \in \IE_* \subset \IE$, and
$R_{\rm out}$ be the infimum of $\mu(e)$ over $e \in \IE \backslash \IE_*$. 
Then $\mu_* = \min(R_{\rm in},R_{\rm out})$ by definition. 
We will show $R_{\rm in} \leq R_{\rm out}$ and hence $R_{\rm in} = \mu_*$. 
For contradiction, suppose $R_{\rm in} > R_{\rm out}$. 
Then there exists $e_o \in \IE \backslash \IE_*$ such that $\mu(e_o) < R_{\rm in}$.
Let us consider the case $e_o>0$. The case $e_o<0$ can be proven similarly.
Since $\IE$ and $\IE_*$ are convex intervals containing $0$, there exists
$e_1 \in (0,e_o)$ such that $(0,e_1) \subset \IE_*$ and $(e_1,e_o) \subset \IE \backslash \IE_*$. 
Then we have
\begin{equation} \label{eq:inequalities} 
R_{\rm in} \leq \varrho_p(e_1) = e_1 < e_o \leq \mu(e_o) < R_{\rm in}. 
\end{equation}
Here, $R_{\rm in} \leq \varrho_p(e_1)$ holds by definition of $R_{\rm in}$ and $(0,e_1) \subset \IE_*$, 
$\varrho_p(e_1) = e_1$ holds since $e_1$ is the upper boundary of $\IE_*$ at which $\varrho_p(e) < |e|$ 
is violated,\footnote{  
In general, there are cases where $\varrho_p(e) > e$ at the upper boundary $e = e_1$ 
of $\IE_*$. In this case, $\varrho_p(e_1) = e_1$ does not hold. 
However, this case occurs only when $e = e_1$ is also the upper boundary of $\IE$, 
and hence $e_o \in \IE \backslash \IE_*$ must be negative. 
That is, whenever we consider the case $e_o > 0$, we must have $\varrho_p(e_1) = e_1$ as claimed.
}
$e_1 < e_o$ and $\mu(e_o) < R_{\rm in}$ hold by definition,
and $e_o \leq \mu(e_o)$ holds due to Lemma~\ref{lem:lbre}.  
However, (\ref{eq:inequalities}) does not hold, 
and hence we conclude $R_{\rm in} \leq R_{\rm out}$ by contradiction. 
\end{proof}

\vspace*{-2mm}

\begin{figure}[t]
\centering
\epsfig{file=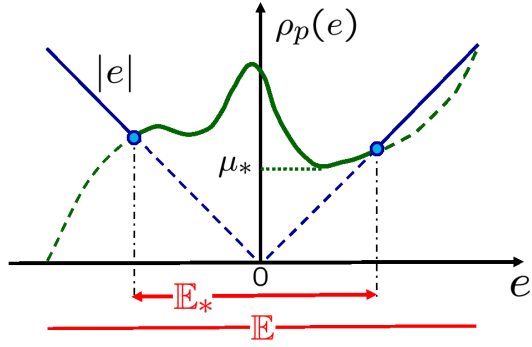,width=70mm}
\vspace*{-1mm}
\caption{Computation for Exact RIR $\mu_*$}
\label{fig:EIM}
\end{figure}

\vspace*{-2mm}

Theorem~\ref{theorem:nonlinear} provides a computable characterization of $\mu_*$ 
at an equilibrium point when conditions (a) and (b) are satisfied.
Figure~\ref{fig:EIM} illustrates the situation related to the proof of 
Theorem~\ref{theorem:nonlinear} by plotting $\varrho_p(e)$ and $|e|$. 
This figure also helps to understand the following 
concrete procedure to calculate the exact RIR $\mu_*$:

\vspace*{-2mm}
\begin{itemize}
\item{Step 1:}
Determine the subset $\IE_* \subset \IE$ defined in Theorem~\ref{theorem:nonlinear} 
by computing $\rhol_p(e)$ for $e \in \IE$. 
\item{Step 2:} 
Check conditions (a) and (b) in Lemma~\ref{lem:re} for $e \in \IE_*$.  
If they are satisfied, then compute the infimum of 
$\varrho_p(e)$ over $\IE_*$ which provides $\mu_*$. 
Otherwise, the infimum gives a lower bound of $\mu_*$. 
\end{itemize}

\vspace*{-2mm}

\section{Applications to Repressilator}
\label{sec:Repressilator}

\vspace*{-2mm}

\subsection{Model of Repressilator}
\label{subsec:ModelRepressilator}

\vspace*{-2mm}

\begin{figure}[h]
\centering
\includegraphics[width=70mm]{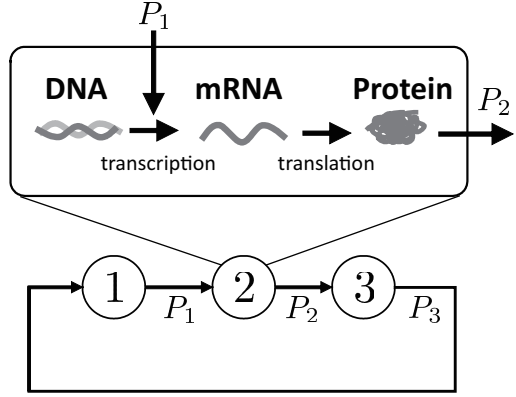}
\caption{Model of the repressilator}
\label{fig:SBS1}
\end{figure}

We consider a class of biomolecular systems in Fig. \ref{fig:SBS1} motivated 
by applications in synthetic biology. 
This system is called repressilator \citep{elowitz:00} and consists of three species of proteins 
P$_i$ $(i=1,2,3)$, each of which is designed to repress the production of another
protein species using the simple cyclic feedback. 
It is known that the repressilator in Fig. \ref{fig:SBS1} has a single equilibrium point \citep{hori:11},  
and thus, destabilization of the equilibrium point leads to oscillatory dynamics of the concentrations of P$_i$, 
given that the trajectories are bounded. 
In the previous work \citep{niederholtmeyer:15}, 
this mechanism was experimentally confirmed \textit{in vitro} by tuning the parameters of synthetic biomolecular oscillators (see \cite{potvin:16} for discussion for \textit{in vivo}).

The nominal dynamical model of the repressilator is given by 
the following ordinary differential equations:
\begin{equation} 
\dot{x}_i(t) = -\alpha_i x_i(t) + \beta_i \psi_i(x_{i-1}(t)), \hs i=1,2,3
\label{system-eq}
\end{equation}
where $x_i(t)$ is the concentration of protein P$_i$, $\alpha_i$ ($>0$) is the degradation rate of P$_i$, 
and $\beta_i$ ($>0$) is the gain of the interactions. 
The index $i$ is defined by modulo 3, implying that $x_0(t) := x_3(t)$.
The function $\psi_i(\cdot)$ is a monotone decreasing static nonlinearity called Hill function \citep{alon:06book} 
that represents the rate of protein production. 
Specifically,
\begin{equation}
\vspace*{-4mm}
\psi_i(x) = \frac{K_i^{\nu_i}}{K_i^{\nu_i}  + x^{\nu_i}}, \hs i=1,2,3
\vspace*{-4mm}
\end{equation}
with a Hill coefficient $\nu_i$ and a Michaelis-Menten constant $K_i$ ($>0$). 

\vspace*{-2mm}

Our theoretical results are verified by the model of a typical experimental setting 
with the parameters chosen based on the experimental data in \citep{niederholtmeyer:15}; 
\begin{equation} \label{parameters} 
\vspace*{-6mm}
\begin{array}{llll} 
 \alpha_1 = 0.4621, \hs & \beta_1 = 138.0, \hs & K_1 = 5.0, \hs & \nu_1 = 3 \\
 \alpha_2 = 0.5545, \hs & \beta_2 = 110.4, \hs & K_2 = 7.5, \hs & \nu_2 = 3 \\
 \alpha_3 = 0.3697, \hs & \beta_3 = 165.6, \hs & K_3 = 2.5, \hs & \nu_3 = 3, 
\end{array} 
\vspace*{-6mm}
\end{equation}
where the units of the parameters $\alpha_i, \beta_i$ and $K_i$ 
are (hr)$^{-1}$, nM $\cdot$ (hr)$^{-1}$ and nM, respectively.

\vspace*{-2mm}

\begin{figure}[tb] 
\begin{minipage}{0.45\textwidth}
\centering 
\epsfig{file=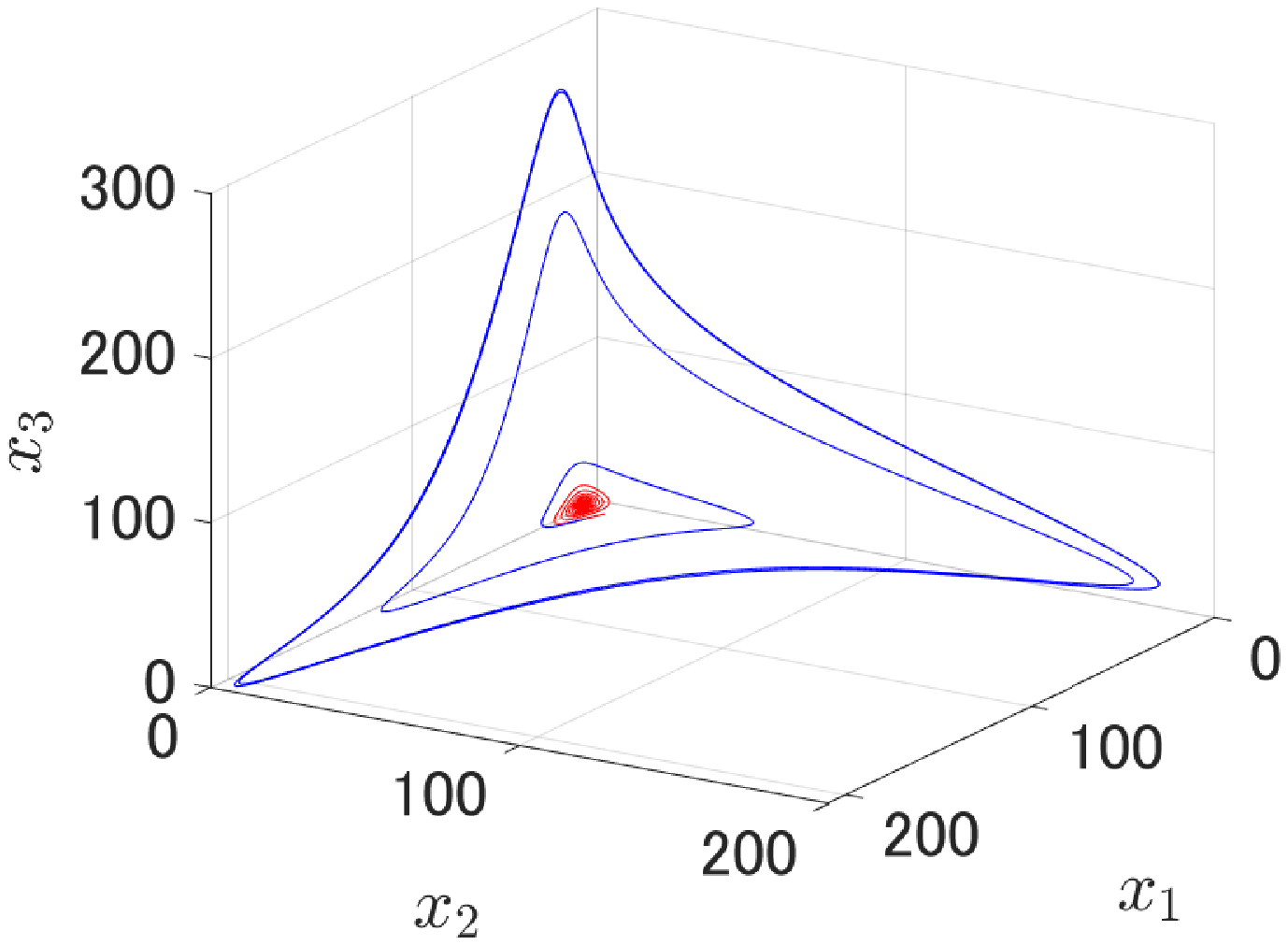,width=60mm} \\
(a) Orbits in $(x_1,x_2,x_3)$ space
\end{minipage} 
\begin{minipage}{0.45\textwidth}
\centering 
\epsfig{file=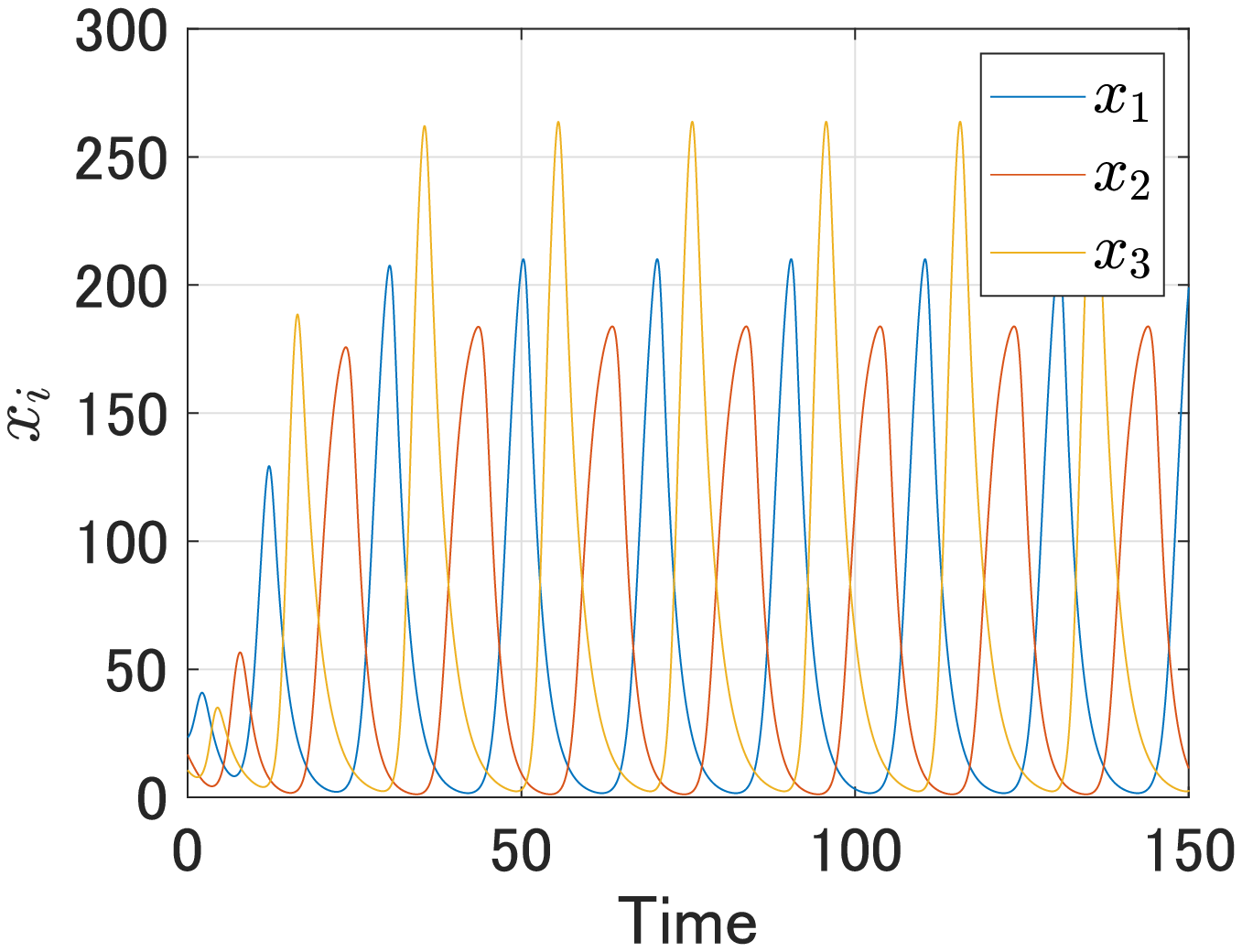,width=60mm} \\
(b) Time responses (nominal)
\end{minipage} 
\caption{Simulations: the repressilator Model}
\label{fig:nominalRep}
\end{figure}

We can readily see by a simple calculation that an equilibrium point exists,
is unique, and  is unstable for this nominal parameter case. 
Consequently, a limit cycle phenomenon can be observed as seen in Figs.~\ref{fig:nominalRep} (a) 
and \ref{fig:nominalRep} (b), 
indicating the blue colored plot of the orbit and the time response, respectively.

\vspace*{-2mm}

In order to investigate the robustness, we assume that there is one perturbation $\delta(s)$, 
which approximately represents the net effect of all the perturbations and the uncertainties in the system. 
This type of assumption has been made in many applications as a crude but effective approximation 
in robust control analysis and design using the small gain condition to avoid increased complexity 
in advanced methods such as $\mu$ synthesis. 
The target system with a multiplicative-type perturbation $\delta(s)$ is then 
represented as 
\vspace*{-2mm}
\begin{equation} \label{rep}
\begin{array}{ll}
\dot x_1=-\alpha_1 x_1 + \beta_1 \psi_1(x_3) + w, &  \hs w = {\hat\delta} z \\ 
\dot x_2=-\alpha_2 x_2 + \beta_2 \psi_2(x_1), &  \\
\dot x_3=-\alpha_3 x_3 + \beta_3 \psi_3(x_2),  &  \\
z = \beta_1 \psi_1(x_3) ,  
\end{array}
\vspace*{-2mm}
\end{equation}
where $\hat\delta$ is the linear operator with the input-output mapping 
specified by stable transfer function $\delta(s)$. 
The purpose of this section is to confirm the effectiveness of the theoretical results in the previous sections 
on the exact RIRs ($\rho_*$) and ($\mu_*$) for the repressilator.   
At an equilibrium, we have 
\begin{equation}
x_i = \frac{\hat\beta_i}{\alpha_i}\psi_i\left( \frac{\hat\beta_{i-1}}{\alpha_{i-1}}
\psi_{i-1} \Big( \frac{\hat\beta_{i-2}}{\alpha_{i-2}}\psi_{i-2}(x_i) \Big) \right) 
\label{eq-eq}
\end{equation}
for $i=1,2,3$, where 
$\hat\beta_1 := (1+e)\beta_1,$, $\hat\beta_2 := \beta_2$, $\hat\beta_3 := \beta_3$, and $e:=\delta(0)$. 
The right-hand side of (\ref{eq-eq}) is a monotonically decreasing function in the positive orthant of $x \in \IR^3$, 
and hence there always exists a unique equilibrium point denoted by $x_e = [x_{e1},x_{e2},x_{e3}]^T$.
Figure~\ref{fig:xe} shows the change of the equilibria due to the change of $e$.

\vspace*{-2mm}

\subsection{Robustness Properties}
\label{subsec:PropertyRepressilator}

\vspace*{-2mm}

\begin{figure}[tb]
\begin{minipage}{0.45\textwidth}
\centering \epsfig{file=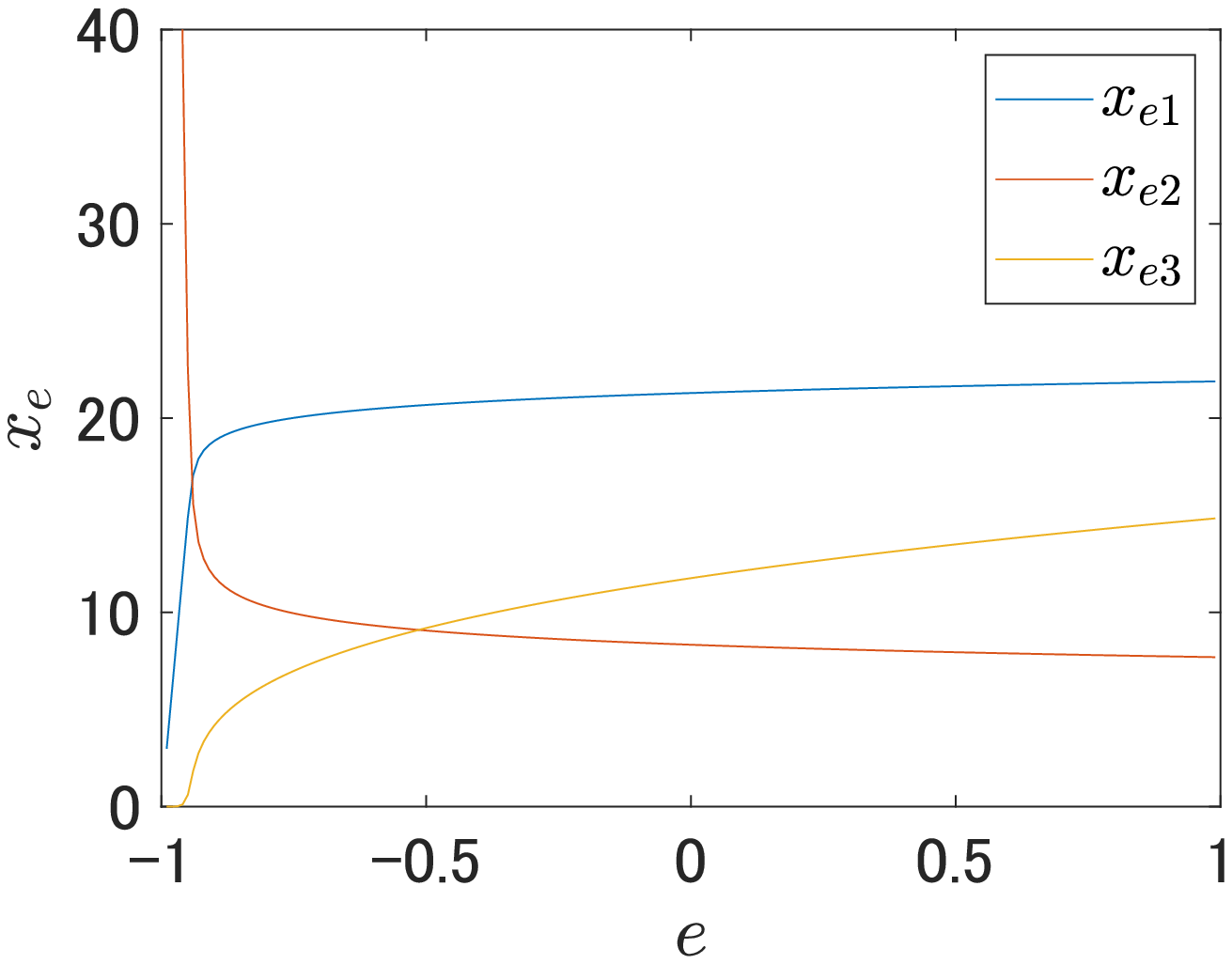,width=60mm} \\
\caption{Equilibrium point} \label{fig:xe} 
\end{minipage} \hfill
\begin{minipage}{0.45\textwidth}
\centering \epsfig{file=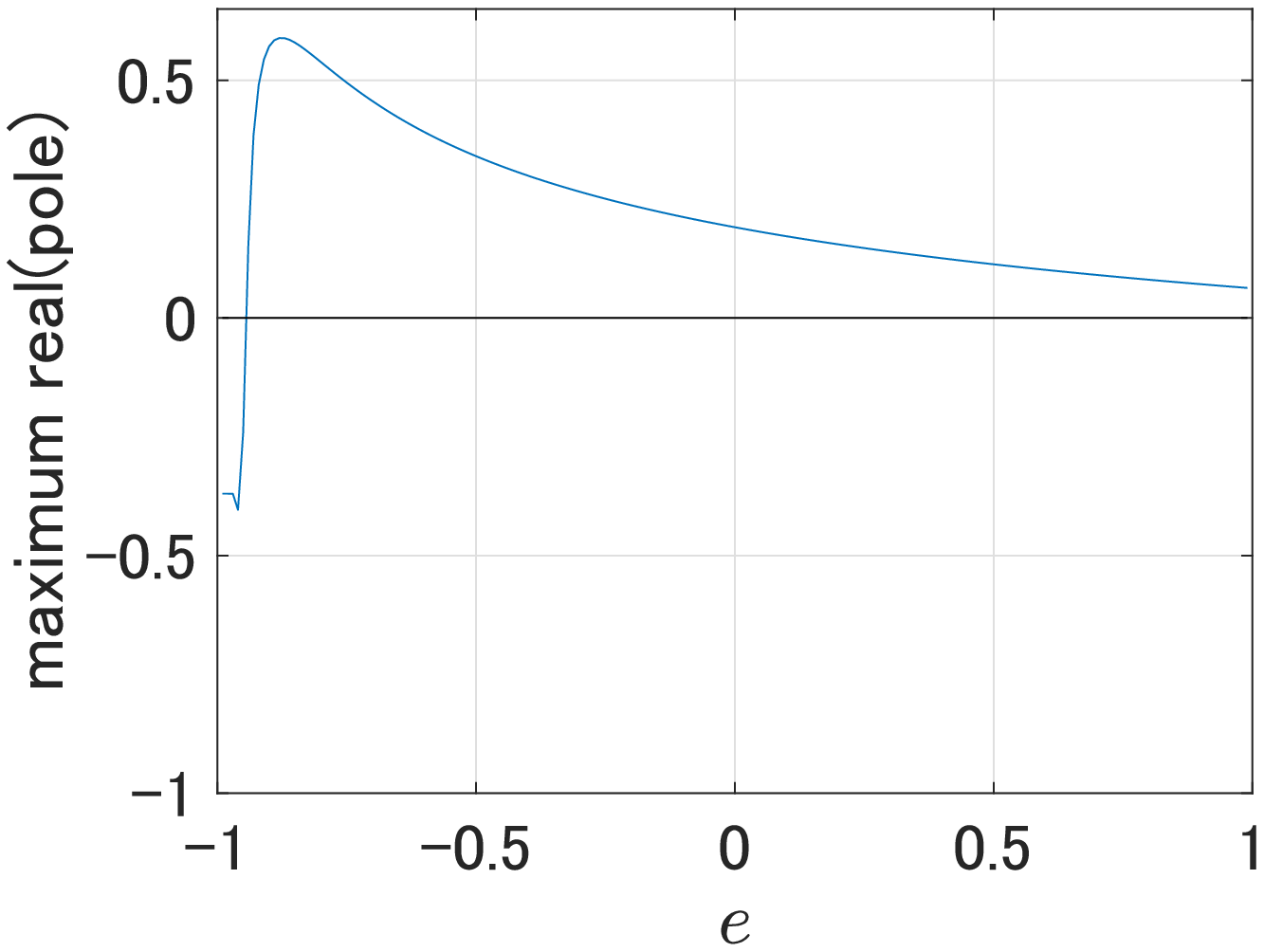,width=60mm}
\caption{Maximum real part of the poles of $g_e(s)$} \label{fig:mrp}
\end{minipage} 
\end{figure}

Let us first show that the repressilator model falls under our analysis framework
and the robust instability radius can be calculated exactly. 
Noting the cyclic structure of the system, the linearization of the system around the equilibrium point is given by
\begin{align}
 \xi=\Big(1+\delta(s)\Big)h_e(s)\xi, \hs \xi:=x-x_e, \hs
\label{cl-eq}
\end{align}
where
\begin{align}
& h_e(s) := \frac{-k}{(s+\alpha_1)(s+\alpha_2)(s+\alpha_3)} \; , 
\label{h3} \\
& k := - \beta_1\beta_2\beta_3 \psi_1'(x_{e3})\psi_2'(x_{e1})\psi_3'(x_{e2})>0,
\end{align}
and the characteristic equation is expressed as 
\begin{equation} \label{che3}
 1=\delta(s)g_e(s), \hs g_e(s)=h_e(s)/(1-h_e(s)).
\end{equation}
It is readily seen that $g_e(s)$ in (\ref{che3}) is represented by  
$g_e(s) = -k/(s^3 + ps^2 + qs + \ell)$, 
where $p := \alpha_1 + \alpha_2 + \alpha_3 > 0$,  
$q := \alpha_1\alpha_2 + \alpha_2\alpha_3 + \alpha_3\alpha_1 > 0$, and 
$\ell := \alpha_1\alpha_2\alpha_3 + k > 0$. 
We now check inequality conditions (\ref{c31}) in Proposition~\ref{coro:3rd} which guarantee Condition 2. 
We can verify that 
\begin{eqnarray*}
\vspace*{-2mm} 
 p^2 - 2q &=& (\alpha_1+\alpha_2+\alpha_3)^2 - 2(\alpha_1\alpha_2 + \alpha_2\alpha_3 + \alpha_3\alpha_1) \\ 
   &=& \alpha_1^2 + \alpha_2^2 + \alpha_3^2 > 0 .
\vspace*{-2mm} 
\end{eqnarray*}
This yields $2p\ell -q^2 > 2p(pq) - q^2 = q \{ 2(p^2 -2q) + 3q\} >0$.  
The remaining condition $\ell > pq$ implies the hyperbolic instability of $g_e(s)$. 
The maximum real part of the poles of $g_e(s)$ for $e \in (-1,1)$ plotted in Fig. ~\ref{fig:mrp} shows that 
$g_e(s)$ is hyperbolically unstable for $e \in \IE_h := (-0.94, 1)$, and it can be confirmed 
that $\ell > pq$ holds for all $e$ in $\IE_h$.  
Hence, we can conclude that (\ref{c31}) holds, which implies that conditions (a) and (b) 
in Lemma~\ref{lem:re} hold as remarked just below Lemma~\ref{lem:re}. 
Consequently, we have $\rho_*(e) = \varrho_p(e)$ for $\delta(0) = e \in \IE_h \subset \IE$, and hence 
we can derive the exact RIR $\mu_*$ by Theorem~\ref{theorem:nonlinear} 
or the procedure presented at the end of Section~\ref{subsec:ExactIM}.

\vspace{-2mm}

For the repressilator, the instability analysis of the equilibrium point 
is in fact sufficient for robustness analysis of the oscillatory behavior.
A precise statement of the result is given as follows.

\vspace{-2mm}

\begin{prop} \label{prop:Yosc}
Consider the repressilator in (\ref{rep}), where, with $i=1,2,3$, all the 
coefficients $\alpha_i$
and $\beta_i$ are positive, nonlinear functions $\psi_i(x)$ are bounded,
continuously differentiable, and satisfy $\psi_i(x)>0$ and $\psi_i'(x)<0$ 
on $x\geq0$, and perturbation $\delta(s)$ is stable. 
There exists a unique equilibrium point in the positive orthant.
Suppose the equilibrium is hyperbolic and unstable,
and the positive orthant
remains to be an invariant set in the presence of the perturbation.
Then, the system is oscillatory in the sense of Yakubovich, i.e., for almost 
all initial states in the positive orthant, the resulting trajectory satisfies
\[
\liminf_{t\rightarrow\infty} x_i(t) < \limsup_{t\rightarrow\infty} x_i(t)
\]
for at least one of the state variables $x_i$.
\end{prop}
\vspace{-2mm}
\begin{proof}
The existence and uniqueness of the equilibrium point follows from (\ref{eq-eq})
as discussed earlier. From (\ref{rep}), the dynamics of $x_1$ is described by
\[
x_1=f_1(s)\psi_1(x_3), \hs f_1(s):=\beta\cdot\frac{1+\delta(s)}{s+\alpha_1}.
\]
Since $\psi_1(x)$ is a bounded continuous function on $x>0$, there is a
scalar $u_1$ such that $|\psi_1(x)|<u_1$ for all $x>0$. Due to the invariance
of the positive orthant, $x_3(t)$ remains positive and hence
$|\psi_1(x_3(t))|<u_1$ holds for all $t\geq0$. 
Since $f_1(s)$ is stable, the effect of the initial condition on $x_1(t)$
will eventually die out and we have $|x_1(t)|\leq\gamma_1u_1$ for sufficiently
large $t$, where $\gamma_1$ is the peak-to-peak gain ($L_1$ norm) of $f_1(s)$.
Similar arguments apply to $x_2$, $x_3$, and the states of $\delta(s)$, and
hence all the trajectories in the positive orthant are ultimately bounded.
The result then follows from Theorem 1 of \citep{pogromsky:99}.
\end{proof}

\vspace{-2mm}

The invariance of the positive orthant after a perturbation is a reasonable 
assumption, given that the variables $x_i$ represent the concentration level 
of proteins. Hence Proposition~\ref{prop:Yosc} basically says that an 
oscillation occurs whenever the equilibrium point is unstable because every 
trajectory repelled from the equilibrium cannot diverge and has to stay in a 
bounded set regardless of the initial condition. Thus, robust instability
of the equilibrium implies persistence of the oscillatory behavior.
This type of analysis has been done for nominal oscillations of central 
pattern generators \citep{futakata:08}, to which our robustness analysis may 
also apply.

\subsection{Illustration by Simulations}  
\label{subsec:Simulations}

\vspace*{-2mm}

We first consider a simple case where the static gain of the perturbation is zero, 
{\it i.e.,} $\delta(0) = e = 0$, to confirm that $\rho_*(0) = \rhol_p(0)$ holds when $g(s)$ satisfies Condition~2 
described in Section \ref{subsec:ExactClass}. 
In this case, the nominal equilibrium  
$x_o=[21.3, 8.34, 11.8]^T$  
remains the same even after the perturbation, {\it i.e.,} $x_e = x_o$.

\vspace*{-2mm}

Using the analytic expression in the proof of Proposition~\ref{coro:3rd},  
the exact RIR for this system is obtained as $\rho_*(0) = \rhol_p(0) = 0.4049$. 
We can confirm that the value is exact as long as no static gain perturbation is allowed by numerical simulations, 
which are not shown here due to the page limitation.

\vspace*{-2mm}

\begin{figure}[t]
\centering \epsfig{file=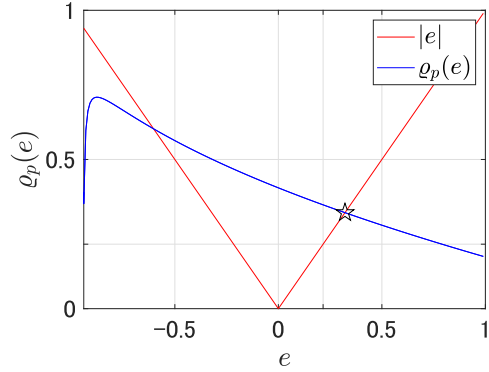,width=70mm} \\
\vspace*{-4mm}
\caption{$\varrho_p(e)$ for $g_e(s)$ (blue curve) and $|e|$ (red lines)}
\label{fig:rhoe}
\end{figure}

We here focus on a more realistic case where the perturbation $\delta(s)$ has 
a non-zero static gain $e:=\delta(0) \neq 0$.
In contrast with the case of $e = 0$, the equilibrium point $x_e$ varies with $e$ as
already seen in Fig.~\ref{fig:xe}. 
The goal is to verify Theorem~\ref{theorem:nonlinear} on the exact RIR $\mu_*$. 
To this end we use the type of plots as shown in Fig.~\ref{fig:EIM}. 
The values of $\varrho_p(e) := 1/\|g_e\|_{L_\infty}$ are plotted as a function of 
$e \in \IE_h = (-0.94, 1)$ 
as seen in Fig.~\ref{fig:rhoe}.
Then we have $|e| < \varrho_p(e)$ when $-0.6027 <e <0.3218$, which defines the set $\IE_*$. 
In this interval, $g_e(s)$ satisfies conditions {\it (a)} and {\it (b)} in Lemma~\ref{lem:re}, 
and hence we conclude $\mu(e) = \varrho_p(e)$. 
The smallest value within this interval $\IE_*$ is $\mu(e) = 0.3218$, which is the exact 
value of the RIR $\mu_*$ at the nominal equilibrium $x_o$ 
since $\mu(e) \geq |e|$ for all $e \in \IE$, and the red lines in Fig.~\ref{fig:rhoe}
give a lower bound on $\mu(e)$.  

\vspace*{-2mm}

\begin{figure}[t]
\begin{minipage}{0.45\textwidth}
\centering \epsfig{file=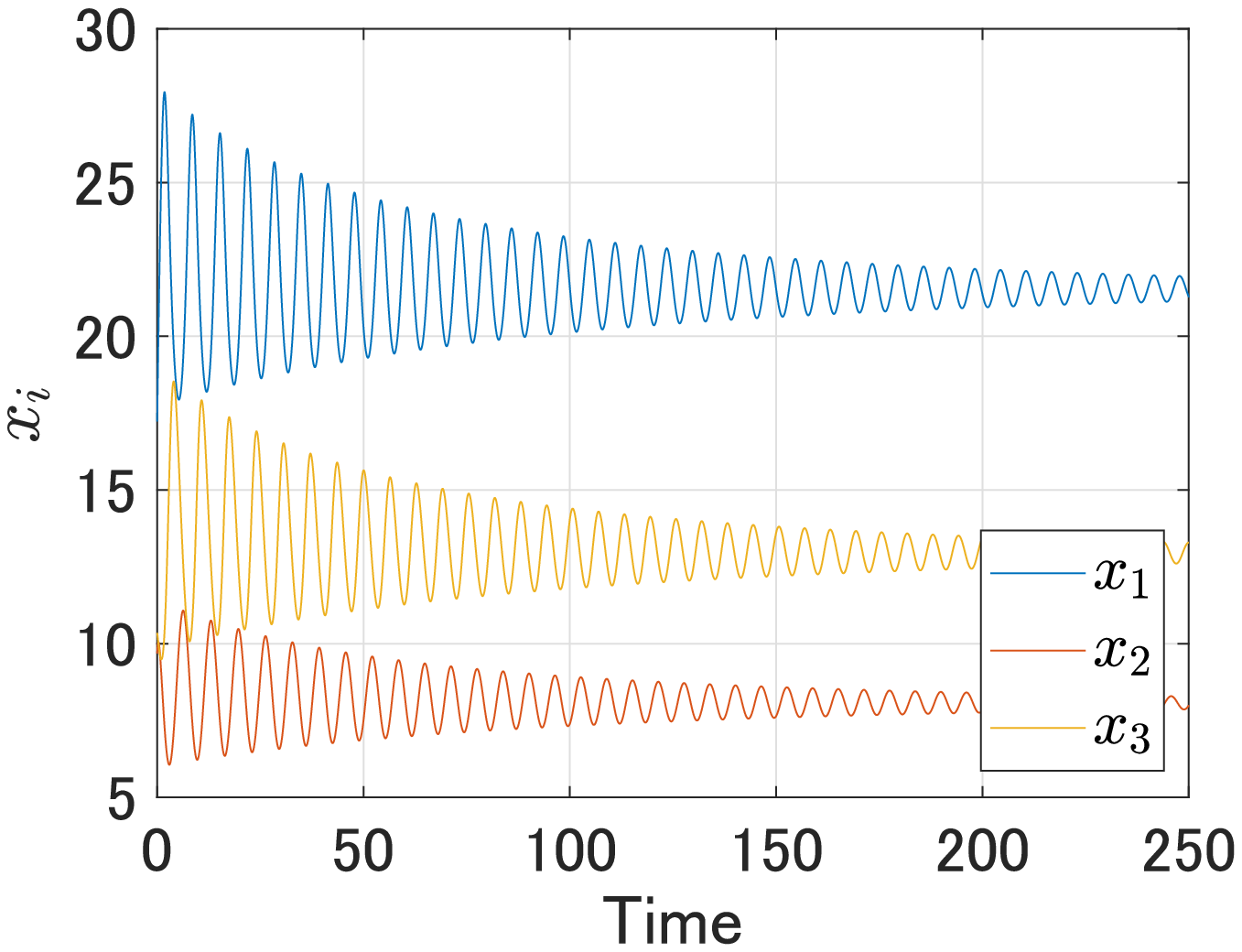,width=60mm} \\
(a) Time responses  (perturbed: $\eps=0.05$)
\end{minipage} \hfill
\begin{minipage}{0.45\textwidth}
\centering \epsfig{file=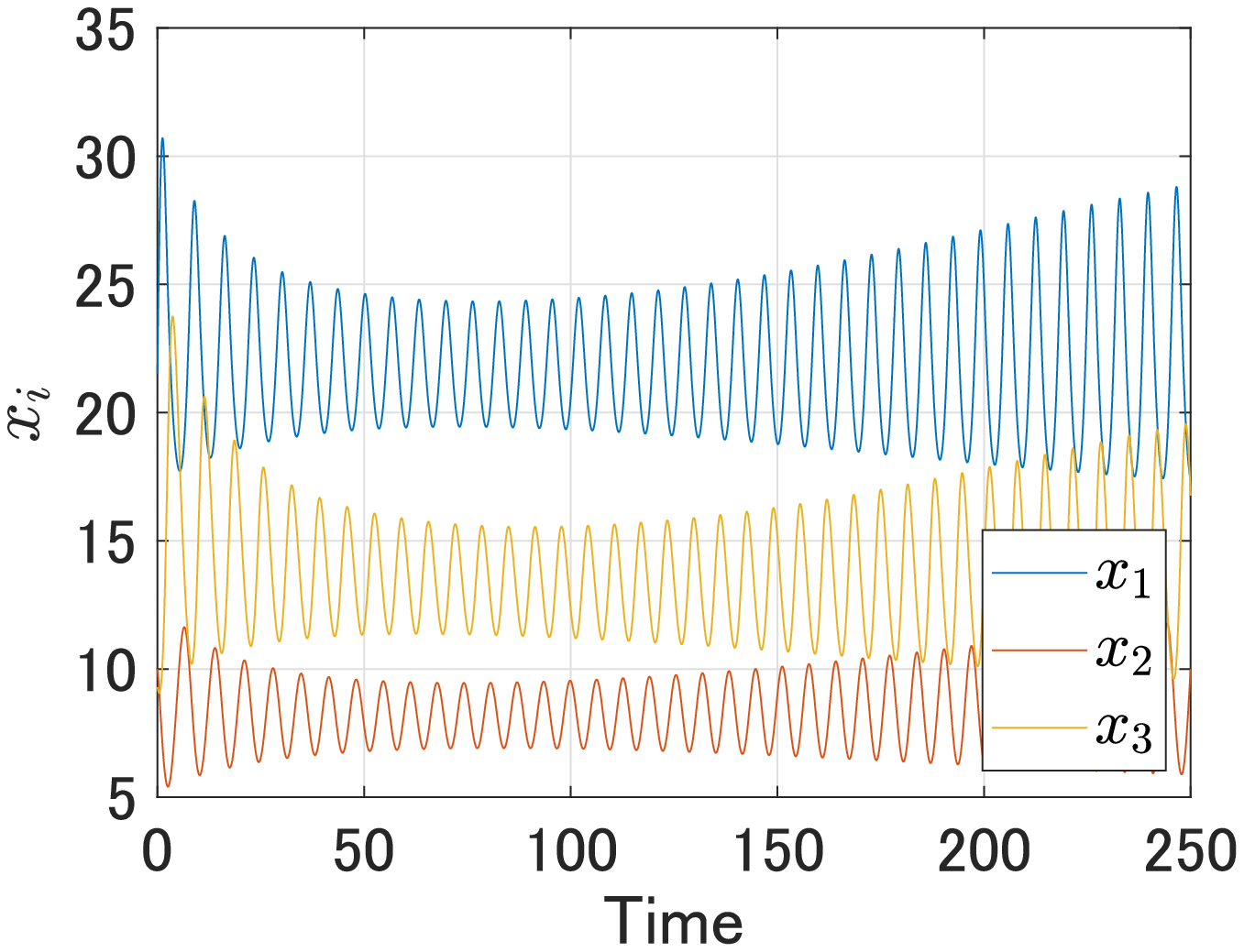,width=60mm} \\
(b) Time responses (perturbed: $\eps=-0.05$) 
\end{minipage} 
\caption{Simulations: Case with change of equilibrium}
\label{fig:RepChaneEP}
\end{figure}

A numerical analysis by simulations depicted in Fig.~\ref{fig:RepChaneEP} reconfirms that 
$\mu(e)$ at $e = 0.3218$ gives the exact RIR $\mu_*$.  
Figure~\ref{fig:RepChaneEP} (a) shows the time response for the case of  
\[
\vspace*{-4mm}
\delta(s) = \frac{s+\xi\gamma}{s+\xi}\cdot\frac{(1+\eps)b(s-a)}{s+a},  \hs
a = 2.253, \hs b=0.3218 
\vspace*{-4mm}
\]
with $\eps = 0.05, \; \gamma = -0.9524 , \; \xi = 0.010$, 
where $\gamma$ is determined by $\gamma = -1/(1+\epsilon)$ to ensure that the high pass filter 
does not change the static gain. 
As illustrated in Figure~\ref{fig:RepChaneEP} (a), this perturbation stabilizes $g_e(s)$ at $e=0.3218$ 
since it satisfies $\delta(0)=0.3218$ and $\|\delta\|_{H_\infty}=0.3379$.
On the other hand, we can observe the maintenance of the periodic oscillation phenomenon 
if we change the sign of $\eps$, meaning that the norm of $\delta(s)$ is smaller than 
$\mu_* = 0.3218$ (See Fig.~\ref{fig:RepChaneEP} (b)). 

\vspace*{-2mm}

\section{Conclusion}
\label{sec:Concl}

\vspace*{-2mm}
  
This paper has provided two main theoretical results on the robust instability 
analysis against stable perturbations.  
One is on the robust instability radius $\rho_*$ for SISO LTI systems, 
and the other is on the robust instability radius $\mu_*$ for parametrized LTI systems.    
The effectiveness of the theoretical results has been illustrated by numerical simulations of the repressilator model. 
This example demonstrated that the theoretical quantitative foundation provided in this paper based on the local 
stability/instability property can lead to a useful tool in the field of synthetic biology. 

\vspace{-2mm}

The classic theory \citep{pogromsky:99} guarantees existence of 
oscillations which may not be periodic under instability of equilibrium points
and ultimate boundedness of trajectories. 
However, periodic orbits may serve better for functional purposes in applications. 
A recent result on the analysis of global nonlinear behaviors based on the concept of $p$-Dominance \citep{forni:19} 
may be useful to guarantee persistence of a limit cycle for a class of systems. 
Toward this direction, our quantitative tool for the instability analysis may be effective 
for checking the $p$-Dominance condition through the spectral splitting.

\vspace{-2mm}

The future work along this research direction includes a characterization of higher order systems for which 
the RIR can be analyzed exactly and its applications to a more general type of biomolecular systems.  

\vspace{-2mm}

{\small 
{\bf Acknowledgments:} \; 
The authors would like to thank Chung-Yao Cao for his valuable comments to improve the paper. 
This work was supported in part by the Ministry of Education, Culture, Sports, Science and Technology in Japan 
through Grant-in-Aid for Scientific Research (A) 21246067 and (B) 18H01464. 
}

\vspace{-3mm}

\bibliographystyle{agsm}
\bibliography{SH_TI_YH_v3}

\vspace*{-3mm}

\appendix
\section{Proof of Proposition~\ref{PropS2}}
\label{subsec:ProofProp3}

\vspace*{-3mm}

Let $\delta_o(s)$ and $g(s)$ be expressed as the ratios of coprime polynomials
$\delta_o(s)=b_o(s)/a_o(s)$ and $g(s)=n(s)/d(s)$, respectively.
Here, $\delta_o(s)$ may be a real constant with $b_o\in\IR$ and $a_o=1$.
Let $b_1(s)$ and $a_1(s)$ be coprime polynomials of the same degree.
For $\delta_1(s):=b_1(s)/a_1(s)$, the characteristic 
polynomial of the perturbed closed-loop system is given by 
$p(s)=\eps q(s)$ with $p(s):= (a_o(s)d(s)-b_o(s)n(s))a_1(s),$ 
and $q(s):=a_o(s)b_1(s)n(s)$. 
Since $\delta_o(s)$ marginally stabilizes $g(s)$ with a simple pole on
the imaginary axis (denote it by $s=j\omega_c$, 
where $\omega_c$ may be zero), the nominal characteristic 
polynomial $p(s)$ takes the 
form $p(s)=(s-j\omega_c)\hat p(s)$, where $\hat p(j\omega_c)\neq0$.
Hence, the characteristic equation can be written as
$s-j\omega_c=\eps r(s), \hs r(s):=q(s)/\hat p(s)$.

\vspace*{-2mm}

We apply the root locus method and focus on the direction of the root locus 
around $s=j\omega_c$ when $\eps$ varies between negative and positive values. 
For small perturbation $|\eps|$, consider the characteristic root 
$\lambda_\eps$ that passes through $j\omega_c$ at $\eps=0$.  
Note that
$\angle(\lambda_\eps-j\omega_c)  =  \angle r(\lambda_\eps)+\angle(\eps)$ 
holds for the phase angles. Taking the limit $\eps\rightarrow0$, 
\[
\vspace*{-2mm}
\angle(\lambda_\eps-j\omega_c) \rightarrow 
\left\{ \begin{array}{ll}
\angle r(j\omega_c), & (\eps\downarrow0), \\
\angle r(j\omega_c)+\pi, & (\eps\uparrow0), 
\end{array} \right.
\vspace*{-2mm}
\]
where the limit is well defined due to $r(j\omega_c)\neq0$, which is 
verified as follows. Note that $r(j\omega_c)=0$ implies 
$a_o(j\omega_c)b_1(j\omega_c)n(j\omega_c)=0$. For a generic $b_1(s)$, 
we have $b_1(j\omega_c)\neq0$. Since $\delta_o(s)$ has no pole on the
imaginary axis, $a_o(j\omega_c)\neq0$. Thus we conclude $n(j\omega_c)=0$.  
Since $s=j\omega_c$ is a pole of the nominal closed-loop system with
$\eps=0$, we have 
$b_o(j\omega_c)n(j\omega_c)=a_o(j\omega_c)d(j\omega_c)=0$.
However, this is a contradiction since $\delta_o(s)$ has no pole on the
imaginary axis and $(n,d)$ are coprime. Thus $r(j\omega_c)$ must be 
nonzero. Now, we may assume, for a generic $\delta_1(s)$, that the real 
part of $r(j\omega_c)$ is nonzero and $\angle r(j\omega_c)\neq\pm\pi/2$. 
Since the phase angle of $\lambda_\eps-j\omega_c$ rotates by $\pi$ when 
passing through $\eps=0$, we see that $\lambda_\eps$ has a negative real 
part when $\eps>0$ or $\eps<0$. If $|\eps|$ is sufficiently small, the
other characteristic roots will stay in the OLHP. 
Thus we conclude the result.

\vspace*{-2mm}

\section{Proof of Proposition~\ref{coro:3rd}}
\label{subsec:ProofsCorollaries}

\vspace*{-2mm}

The proof of the first part is easy. 
The requirement of the odd number of the ORHP poles of $g(s)$ is equivalent to $\ell<0$.
For a constant $\delta$, the closed-loop characteristic equation is given by
$s^3 + ps^2 + (q-\delta \zeta)s + (\ell + k\delta) = 0$.
When $\delta = -\ell/k$, one root is at the origin, and the remaining two
roots are in the OLHP if and only if $p>0$ and $q - \delta\zeta = q + \zeta \ell/k > 0$.
Thus we have (\ref{c22}).

\vspace*{-2mm}

Now we define  $\psi(s) := 1/g(s)$ for the proof of the second part. 
Letting $\Omega := \omega^2$, $|\psi(j\omega)|^2$ is given by
\begin{equation} \label{eq:3rd_F}
\vspace*{-2mm}
 F(\Omega) := (\Omega^3 + f_2\Omega^2 - f_1\Omega + f_0)/(\zeta^2\Omega + k^2) , 
\vspace*{-2mm}
\end{equation}
where $f_2 := p^2 - 2q$, $f_1 := 2p\ell - q^2 > 0$, and $f_0 := \ell^2 > 0$. 
We now seek the critical frequency $\omega_p$ which provides the minimum of $F(\Omega)$ by 
calculating $dF(\Omega)/d\Omega$. 
It is seen that $dF(\Omega)/d\Omega = 0$ is equivalent to  
\begin{eqnarray} \label{eq:3rd_H}
\vspace*{-2mm}
  H(\Omega) &:=& 2\zeta^2\Omega^3 + (\zeta^2 f_2 + 3k^2)\Omega^2 \nonumber \\ 
  && + 2k^2 f_2 \Omega - (\zeta^2 f_0 + k^2 f_1) = 0 .  
\vspace*{-2mm}
\end{eqnarray}
We show that $H(\Omega)$ has a unique positive solution $\Omega_p$ which corresponds to 
the critical frequency $\omega_p^2 \neq 0$. 
First note that $f_0 > 0$ and $f_1>0$ imply $H(0) <0$. 
$dH(\Omega)/d\Omega$ is positive for all $\Omega > 0$ if $f_2 \geq 0$ and 
$dH(\Omega)/d\Omega$ at $\Omega = 0$ is negative if $f_2 < 0$ .
These facts conclude that $H(\Omega)$ has a unique positive solution $\Omega_p$. 

Hereafter we will show that $\delta_o(s)$ defined by (\ref{do}) marginally stabilizes $g(s)$, 
which means that
the characteristic equation $1 - \delta_o(s)g(s) = 0$, or  
\begin{equation}
\vspace*{-2mm}
   (s^3 + ps^2 + qs + \ell)(s+a) -  b(\zeta s -k)(s-a) = 0 
\vspace*{-2mm}
\label{eq:3rd_D1}
\end{equation}
has a form of 
\begin{equation} \label{eq:3rd_D2}
\vspace*{-2mm}
 (s^2 + \Omega_p)(s^2 + \sigma_1 s + \sigma_0) = 0  
\vspace*{-2mm}
\end{equation}
for a certain positive parameters $\sigma_1$ and $\sigma_0$.  
Comparing the coefficients of (\ref{eq:3rd_D1}) and (\ref{eq:3rd_D2}), we have
\begin{eqnarray}\label{eq:3rd_Axb}
\vspace*{-2mm}
&&
 {\bf A}{\bf x} = {\bf b} , \hs 
 {\bf x} := \begin{mat}{ccc} a  & \sigma_1 & \sigma_0 \end{mat}^T ,  \\
&& 
{\bf A} : = 
\begin{mat}{ccc}
-1 & 1 & 0 \\ -p & 0 & 1 \\  -(q+\zeta b) & \Omega_p & 0 \\ kb - \ell & 0 & \Omega_p
\end{mat}, \hs 
{\bf b} := 
\begin{mat}{c} p  \\ q - \zeta b - \Omega_p \\  \ell + kb \\ 0 \end{mat}.  \nonumber 
\vspace*{-2mm}
\end{eqnarray}
It is clear that $\mbox{rank} \; {\bf A} = 3$ and that ${\bf A_b} := \begin{mat}{cc} {\bf A} & {\bf b} \end{mat}$ 
is singular because the determinant of ${\bf A_b}$ is equal to 
$(\zeta^2\Omega_p + k^2)(F(\Omega_p) - b^2) = 0$. 
This guarantees the existence of the unique solution of (\ref{eq:3rd_Axb}). 

\vspace*{-2mm}

Consequently, the remaining step of the proof is to show the positivity of the solution 
${\bf x}$, {\it i.e.}, $a>0$,  $\sigma_1>0$, and $\sigma_0>0$. 
Note that $b$ defined in (\ref{eq:ab}) for $\omega_c = \omega_p$ satisfies 
\[
\vspace*{-2mm}
|b| = 1/\max_{\omega \neq 0} |g(j\omega)| = \min_{\omega \neq 0} |\psi(j\omega)|  
 <  \ell/|k| = |\psi(0)| 
\vspace*{-2mm}
\]
and that $a$ can always be chosen to be positive, depending on 
the sign of $b$ as described in (\ref{eq:ab}).

Under the assumptions of $a>0$ and $|kb| < \ell$ with $p>0$, 
the 1st and the 4th rows in (\ref{eq:3rd_Axb}) yield 
$\sigma_1 = a + p > 0$ and $\sigma_0 = (\ell - kb)a/\Omega_p > 0$, respectively. 
This completes the proof of the second part. 

\vspace*{-2mm}

\section{Proofs of Lemmas~\ref{lem:lbre}, \ref{lem:hpf}, and \ref{lem:re}}
\label{subsec:ProofsLemmas}

\vspace*{-2mm}

\noindent
{\bf $\bullet$ Proof of Lemma~\ref{lem:lbre}}

\vspace{-2mm}

The first condition $\mu(e) \geq |e|$ follows from the definition of $\mu(e)$ 
in (\ref{re}) because $\|\delta\|_{H_\infty}\geq |\delta(0)|=|e|$ for $\delta(s) \in \Delta_e$. 
In the second condition, we obtain $\mu(e) \geq \rho_*(e)$ by inspection of (\ref{rhoe}).

\noindent
{\bf $\bullet$ Proof of Lemma~\ref{lem:hpf}}

\vspace{-2mm}

The characteristic equation in (\ref{chefl}) can be rewritten as
\begin{equation} \label{chexiL}
1+\frac{\xi}{s}L(s)=0, \hs 
L(s):=\frac{1-\gamma \ell(s)}{1-\ell(s)}.
\end{equation}
Note that $L(s)$ is stable since $1=\ell(s)$ implies $\Re(s)<0$.
We claim that $L(s)$ has an even number of zeros (including multiplicities) in the ORHP and
no zeros on the imaginary axis. This is easy to see for the case $\gamma=0$
because the zeros of $L(s)$ coincide with the poles of $\ell(s)$.
When $\gamma\neq0$, by the small gain condition in (\ref{sgc}), 
$1=\gamma_o\ell(s)$ has no roots on the imaginary
axis for all $\gamma_o$ such that $|\gamma_o|\le|\gamma|$. 
Since $\ell(s)$ has an even number of poles in the ORHP and no poles on the
imaginary axis, there are even number of roots of 
$1=\gamma_o\ell(s)$ in the ORHP when $|\gamma_o|$ is nonzero and
sufficiently small. As $|\gamma_o|$ increases to $|\gamma|$, none of the
roots of $1=\gamma_o\ell(s)$ can go across the imaginary axis, and hence
$1=\gamma\ell(s)$ has an even number of ORHP roots.
Thus $L(s)$ has no zero on the imaginary axis and an even number of zeros in 
the ORHP. Then the root locus shows that all the roots of the characteristic
equation in (\ref{chexiL}) are in the OLHP for sufficiently small $\xi>0$.

\noindent
{\bf $\bullet$ Proof of Lemma~\ref{lem:re}}

\vspace{-2mm}

Let $\delta_e(s)$ be a transfer function as described in {\it (a)}. 
Then a slight perturbation of $\delta_e(s)$ can 
stabilize $g_e(s)$ as shown in Proposition~\ref{PropS2}. 
That is, for an arbitrarily small $\eps>0$, there exists a stable 
transfer function $\tilde\delta_e(s)$ that stabilizes $g_e(s)$ and satisfies 
$\|\delta_e-\tilde\delta_e\|_{H_\infty}<\eps$.
Now, the static gain of the perturbation $\tilde\delta_e(s)$ is 
approximately given by $\tilde\delta_e(0)\cong\delta_e(0) = \varrho_p(e)$, and
hence this perturbation may not belong to $\Delta_e$. Let the static gain 
of the perturbation be adjusted by a high pass filter
\[
\vspace*{-2mm}
\delta(s):=f(s)\tilde\delta_e(s), \hs
f(s):=\frac{s+\xi\gamma}{s+\xi}, \hs \gamma:=\frac{e}{\tilde\delta_e(0)},
\vspace*{-2mm}
\]
so that $\delta(0)=e$. We will show that $\delta(s)$ with sufficiently 
small $\xi>0$ stabilizes $g_e(s)$ and hence $\delta(s)\in\Delta_e$, using
Lemma~\ref{lem:hpf} with $\ell(s):=\tilde\delta_e(s)g_e(s)$,
where the characteristic equation $1=\delta(s)g_e(s)$ is given by (\ref{chefl}). 
First note that $\ell(s)$ has an even number of poles in the 
ORHP and no poles on the imaginary axis because $g_e(s)$ is hyperbolic and 
satisfies condition {\it (b)}, and $\tilde\delta_e(s)$ is stable. 
Next, all the roots of $1=\ell(s)$ are in the OLHP since $\tilde\delta_e(s)$ 
stabilizes $g_e(s)$. Also note that $|\gamma|<1=\|f\|_{H_\infty}$ for all
$\xi>0$ since $|\tilde\delta_e(0)| \cong \varrho_p(e)$ and $|e| < \varrho_p(e)$, and 
hence $\|\delta\|_{H_\infty} \cong \varrho_p(e)$. Moreover,
we have $\|\gamma\ell\|_{L_\infty}<1$ because
\begin{equation} \label{gamdelbnd}
\vspace*{-2mm}
\|\gamma\tilde\delta_e\|_{H_\infty}=
|e|\cdot \frac{\|\tilde\delta_e\|_{H_\infty}}{|\tilde\delta_e(0)|} < \varrho_p(e)
\vspace*{-2mm}
\end{equation}
holds, where the inequality follows from the fact that 
$\|\tilde\delta_e\|_{H_\infty}/|\tilde\delta_e(0)|$ is arbitrarily close to 
$1$ because $\delta_e(s)$ is all pass and 
$\|\tilde\delta_e-\delta_e\|_{H_\infty}$ is arbitrarily small.
Thus, all the conditions
in Lemma~\ref{lem:hpf} are satisfied and we conclude that
$\delta(s)$ with sufficiently small $\xi>0$ stabilizes $g_e(s)$ 
and hence $\delta(s)\in\Delta_e$. The proof is now complete by noting 
that $\mu(e) \leq \|\delta\|_{H_\infty} \cong \varrho_p(e)$ due to the preceding
argument and $\varrho_p(e) \leq \mu(e)$ due to Lemma~\ref{lem:lbre}.
\end{document}